\documentclass[twocolumn]{aastex61}

\newcommand{\Rnum}[1]{\uppercase\expandafter{\romannumeral #1\relax}}

\received{---}
\revised{---}
\accepted{---}
\submitjournal{PASP}

\shorttitle{Characterization of CANICA}
\shortauthors{Devaraj et al.}

\begin{document}

\title{Characterization and Performance of the Cananea Near-Infrared Camera (CANICA)}

\correspondingauthor{Devaraj Rangaswamy}
\email{dev2future@yahoo.com}

\author{Devaraj R.}
\affiliation{Instituto Nacional de Astrof\'isica, \'Optica y Electr\'onica, \\ Luis Enrique Erro \# 1, Tonantzintla, Puebla - 72840, M\'exico}

\author{Y.D. Mayya}
\affiliation{Instituto Nacional de Astrof\'isica, \'Optica y Electr\'onica, \\ Luis Enrique Erro \# 1, Tonantzintla, Puebla - 72840, M\'exico}

\author{L. Carrasco}
\affiliation{Instituto Nacional de Astrof\'isica, \'Optica y Electr\'onica, \\ Luis Enrique Erro \# 1, Tonantzintla, Puebla - 72840, M\'exico}

\author{A. Luna}
\affiliation{Instituto Nacional de Astrof\'isica, \'Optica y Electr\'onica, \\ Luis Enrique Erro \# 1, Tonantzintla, Puebla - 72840, M\'exico}



\begin{abstract}

We present details of characterization and imaging performance of the Cananea Near-infrared camera (CANICA) at the $2.1\,\mathrm{m}$ telescope of the Guillermo Haro Astrophysical Observatory (OAGH) located in Cananea, Sonora, M\'exico. CANICA has a HAWAII array with a HgCdTe detector of 1024 $\times$ 1024 pixels covering a field of view of $5.5 \times 5.5\,\mathrm{arcmin^2}$ with a plate scale of $0.32\,\mathrm{arcsec/pixel}$. The camera characterization involved measuring key detector parameters: conversion gain, dark current, readout noise, and linearity. The pixels in the detector have a full-well-depth of $100,000\,\mathrm{e^-}$ with the conversion gain measured to be $5.8\,\mathrm{e^-/ADU}$. The time-dependent dark current was estimated to be $1.2\,\mathrm{e^-/sec}$. Readout noise for correlated double sampled (CDS) technique was measured to be $30\,\mathrm{ e^-/pixel}$. The detector shows 10\% non-linearity close to the full-well-depth. The non-linearity was corrected within 1\% levels for the CDS images. Full-field imaging performance was evaluated by measuring the point spread function, zeropoints, throughput, and limiting magnitude. The average zeropoint value in each filter are $J$ = 20.52, $H$ = 20.63, and $K$ = 20.23. The saturation limit of the detector is about sixth magnitude in all the primary broadbands. CANICA on the $2.1\,\mathrm{m}$ OAGH telescope reaches background-limited magnitudes of $J$ = 18.5, $H$ = 17.6, and $K$ = 16.0 for a signal-to-noise ratio of 10 with an integration time of $900\,\mathrm{s}$.

\end{abstract}

\keywords{Instrumentation: detectors, Methods: data analysis, Techniques: photometric}



\section{Introduction}



Modern astronomy is driven by large collaborative projects that make use of available resources in an efficient way possible. Such collaborations have led to innovative instruments resulting in extraordinary use of $2-4\,\mathrm{m}$ class
telescopes (e.g. SDSS \citep{york00}, 2MASS \citep{skrut06}, VISTA \citep{minniti10}, CALIFA \citep{sanchez12}, etc.). However, there still exist an innumerable $2\,\mathrm{m}$ class of telescopes that are driven by individual small projects. To keep in pace with the current trends, the existing suite of instruments in these telescopes need to be evaluated and upgraded to be relevant in this context. Near-infrared imaging instruments offer a great opportunity in this direction.

As such, the Cananea Near-infrared camera (CANICA) \citep{car17} is one of the main instruments commissioned at the $2.1\,\mathrm{m}$ telescope of the Guillermo Haro Astrophysical Observatory (OAGH) located in Cananea, Sonora, M\'exico. CANICA has been in operation since 2002 and has carried out a number of astrophysical studies in the near-infrared \citep[e.g.][]{mayya05,mayya06,rodlh11}. A new instrument called POLICAN \citep{dev15,dev17} was implemented in 2012 alongside CANICA for carrying out linear polarimetric studies. In the fall of 2016, the primary mirror of the telescope was re-aluminized to enhance its reflective capabilities. Additionally, upgrades in the telescope console were made with new software developments for the camera operation. 

The recent upgrades with the telescope and the instrument enables CANICA to expand its science goals varying from Galactic star forming regions to extragalactic sources such as active galactic nuclei (AGNs) \citep[e.g.][]{carp01, sugitani02, bonning12}. The polarimetric capabilities of POLICAN permit observations to study scattered polarization and magnetic fields in the interstellar medium \citep[e.g.][]{tamura06, chapman11, clem12}. Essential to all these studies is a full understanding of the capabilities and limitations of CANICA in its current state. Because CANICA is a customized re-imaging camera, its performance needed to be evaluated on the telescope as a fully functional unit. The characteristic lab values of HgCdTe detector with 1024 $\times$ 1024 pixels have to be re-measured to establish the true behavior during camera operation. CANICA offers a field of view (FOV) of $5.5 \times 5.5\,\mathrm{arcmin^2}$ with a plate scale of $0.32\,\mathrm{arcsec/pixel}$. This feature allows observational study of both point and extended sources, for which the image quality, point spread function (PSF), and photometric zeropoints need to be determined. Equally important is knowledge of how these quantities vary across the FOV and with different filter configurations.

The paper presents a two-part description of CANICA's characteristics and performance. The first part describes measurement of key detector parameters such as conversion gain, dark current, readout noise, and linearity, which rely on images taken from dome flats, darks, and BIAS (see Section~\ref{dchar}). The second part describes the imaging performance of CANICA on the telescope: detailing the PSF, zeropoints, throughput, and limiting magnitude, all measured from a combination of multiple observations on the sky. Bridging the detector characteristics and imaging performance is a brief description of the observing scheme and image reduction process.

\begin{figure}[ht!]
\epsscale{1.2}
\plotone{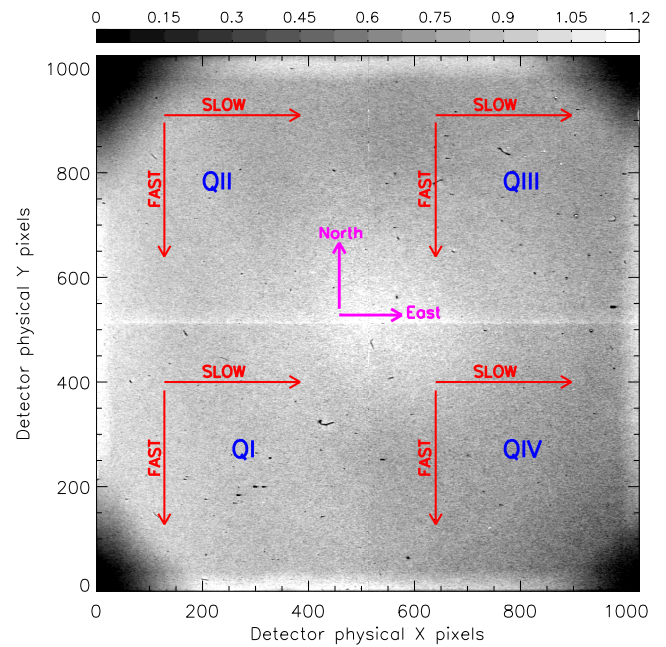}
 \caption{Normalized dome flat in $H$-band marked with each of the four quadrants of the detector. The dark corners in the image are formed due to vignetting. The direction of readout of the shift registers is shown with their fast and slow axis. The equatorial cardinal directions are marked with respect to the detector physical pixels.}
\label{fig1}
\end{figure}

\begin{deluxetable*}{lccc}
\tablecolumns{4}
\tablewidth{0pc}
\tablecaption{CANICA characteristics.\label{tbl-1}}
\tablehead{\colhead{Quantity} & \colhead{Value}  & \colhead{Unit}  & \colhead{Description}}
\startdata
Detector material 			& HgCdTe 				& 		 		& HAWAII array 	\\	
Detector format 			 	& 1024 $\times$ 1024 	& pixels	 		& 4 Quadrants \\	
Pixel size 					& 18.5 					& $\mu$m	 		& Square pixels \\
\hline	
Spectral Range 				& 0.85$-$2.40		 	& $\mu$m 		& $>$85\% filter transmission \\
Focal ratio					& \textit{f/6}			& 			 	& Input beam \textit{f/12} \\
Average FWHM PSF				& 1.5					& arcsec			& In $H$-band \\
Plate scale 					& 0.32 					& arcsec/pixel 	& On the detector \\
Full field of view 			& $5.5\times5.5$ 		& arcmin$^{2}$ 	& Unvignetted FOV $4\times4$\\
Operating temperature 		& 77$-$80 				& K		 		& Liquid nitrogen cooled \\
\hline
Full-well capacity	 		& 100,000 				& e$^-$			& Lab value\\
Saturation limit				& 17,200					& ADUs			& Measured value\\
Conversion gain 				& 5.8	 				& e$^-$/ADU 		& Measured value\\
Dark current			 		& 1.2 					& e$^-$/sec 		& Measured value\\
Readout noise 				& 30 					& e$^-$/pixel 	& Measured value for CDS readout\\
Bias gate voltage			& 3.62					& V				& Kept between $3.3 - 3.8\,\mathrm{V}$\\
Quantum efficiency	 		& $>$55					& \%				& for 99.5\% of the pixels \\
CDS readout time				& 1 						& sec 			& Sum of first and second read times. \\
Linear response		 		& 10\% non-linear 		&				& close to the full-well-depth\\
\enddata
\end{deluxetable*}

\section{Detector Characteristics} \label{dchar}

CANICA offers near-infrared (NIR) imaging capabilities at multiple bands including $J(1.24\,\mathrm{\mu m})$, $H(1.63\,\mathrm{\mu m})$, and $K^{'} (2.12\,\mathrm{\mu m})$ broadbands. The details on CANIAC design, construction, and filter configurations are described in \citet{car17}. CANICA is based on the HAWAII focal plane array \citep{hodapp96} with a HgCdTe detector capable of efficiently detecting light from 0.85 to $2.40\,\mathrm{\mu m}$. The HAWAII array is designed to have mean quantum efficiency (QE) greater than 50\% in the NIR spectral range. More than 99.5\% of the array has full response in all the primary broadbands. The detector array has around 0.2\% bad pixels; most of them are single pixels spread throughout the field with a few seen as clumps. The array consists of four independent quadrants of $512 \times 512$ pixels structured to carry out simultaneous readouts having four outputs each. Basic operation is carried out by six CMOS-level clocks (\textit{Pixel, Lsync, Line, Fsync, ResetB, Read}), two $5\,\mathrm{V}$ power supplies (one analog and one digital) and two DC bias voltages (one fixed and one variable). 

The clock signals are managed by the electronic boards in the CCD controller acquired from Astronomical Research Cameras Inc., USA, originally developed by San Diego State University (SDSU). The SDSU CCD controller synchronizes the clock signals at a rate of $50\,\mathrm{MHz}$, which are fed to the detector array through the pre-amplifier circuit. Each quadrant in the array consists of two digital shift registers: a horizontal and a vertical register, for addressing the pixel readout. The output is obtained when the horizontal register is clocked in the slow direction by clocks \textit{Pixel} and \textit{Lsync}, while the vertical register is clocked in the fast direction by clocks \textit{Line} and \textit{Fsync}. The image in Figure~\ref{fig1} shows a dome flat marked with each of the quadrants having various readout directions of the shift register.

The readout of the HAWAII array in CANICA is carried out by correlated double sampling clocking method, or reset-read-read mode. With this readout method, the clock signals are applied such that the array is reset, read, allowed to integrate, and re-read with the difference between the first and second reads recorded and named as a correlated double sampled (CDS) image. In the CANICA readout structure, the first read is known as the BIAS image and the second read is known as the RAW image, with the CDS image being the difference of RAW and BIAS. The DC bias voltages control the bias level and are operated between $3.3 - 3.8\,\mathrm{V}$. The operational readout time for a CDS image with CANICA is $1\,\mathrm{s}$, with an additional time of $7\,\mathrm{s}$ for image delivery and storage. The BIAS and CDS images are stored for all observations. The detector characteristics and various camera parameters are summarized in the Table~\ref{tbl-1}. 

\begin{figure*}[!ht]
\epsscale{1.15}
\plotone{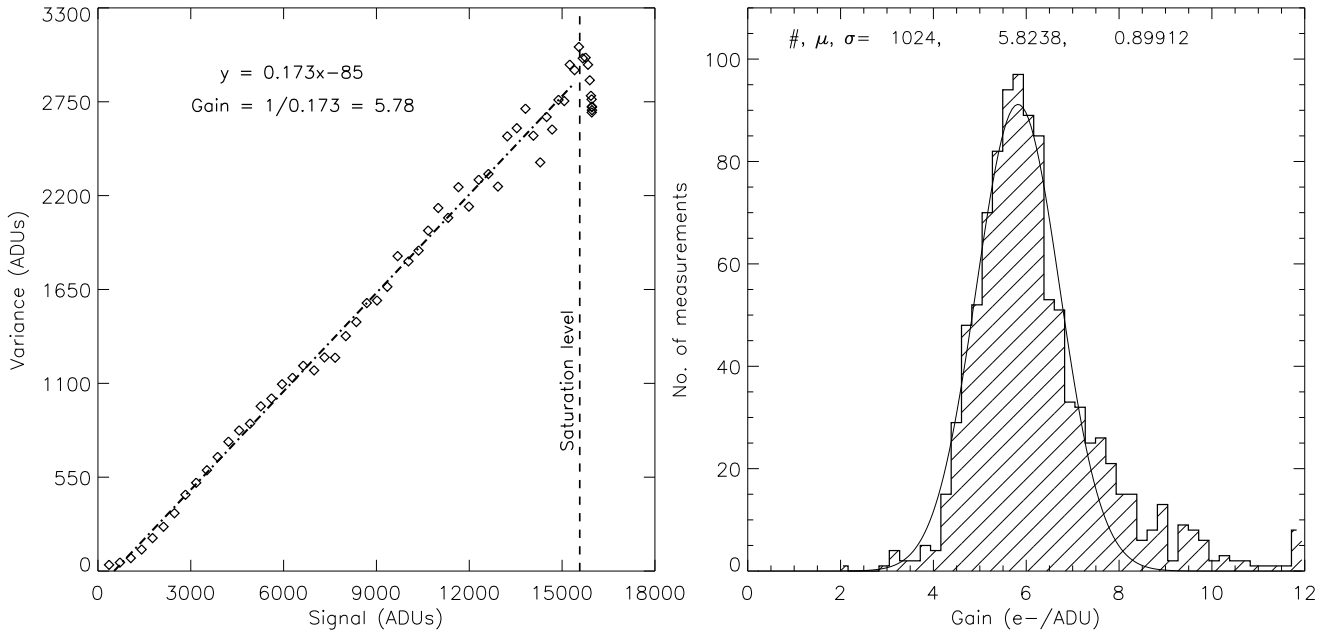}
 \caption{The \textit{left panel} shows data points representing the photon transfer curve for box region of 32 $\times$ 32 pixels in quadrant Q\Rnum{1}. The values of variance and signal are obtained from a series of dome illuminated flats at different exposure time, stepped in increasing order. The data points are fitted with a first-order polynomial up to the saturation level. The reciprocal of the slope of the fit represents the gain value. The \textit{right panel} shows histogram distribution of gain measurements obtained for multiple box regions across the entire detector array. A Gaussian is fitted to the distribution and its peak represents the mean gain value of CANICA.}
\label{fig2}
\end{figure*}

\subsection{Conversion gain}

In a camera system, the conversion of detected photoelectrons to digital units is linearly related and is produced by capacitor (V/$e^{-}$), output amplifier (V/V), and ADC (ADUs/V) \citep{jane01}. Hence, the total conversion gain of the camera is a combination of the above systems and is expressed in $e^{-}$/ADU. The photon transfer technique is widely used to measure number of detector parameters in absolute terms. Gain measurement can be performed using the photon transfer curve by obtaining the slope of a variance-signal plot. 

In theory, the equation for variance-signal plot can be written as \citep[see Section~9.1]{mclean08}:
\begin{equation} \label{eqn1}
N_{c}^2 = \frac{1}{g}(S_{c}) + R_{c}^2 \quad 
\end{equation}
where $S_{c}$, $N_{c}$, and $R_{c}$ are signal, noise, and readout noise, respectively, in counts (ADUs). $g$ is the conversion gain.

This represents equation of a straight line with a slope of $1/g$. Equation~\ref{eqn1} can be used as a good proxy to obtain the conversion gain when the pixel-to-pixel variations are corrected by flat fielding.

Gain measurement of CANICA was carried out using dome illuminated flats taken in $H$-band. Multiple sets of 10 dome flats were obtained for exposure times ranging from $1\,\mathrm{s}$ to $60\,\mathrm{s}$ in increments of $1\,\mathrm{s}$. The 10 flats for each exposure were averaged to obtain the mean flat image with a high signal-to-noise ratio (S/N). The mean flats were first dark subtracted and corrected for pixel-to-pixel variations by dividing with a normalized flat. Next, for each mean flat, the value of signal (mean counts) and its standard deviation were measured for a small box region of 32 $\times$ 32 pixels. The variance was obtained as the square of standard deviation and was corrected for noise increase due to flat-fielding errors (see Appendix~\ref{appA}). The measured values from all the mean flat images were examined to produce the plot of variance against signal. The plot of variance against signal showed a strong rollover after certain linear increase. This value is the saturation level of the detector and was determined to be $17,200\,\mathrm{ADUs}$. A first-order polynomial was fit to the plot up to the saturation level to obtain the slope. The gain value was then measured as the reciprocal of the slope of the fit.

\begin{figure*}[!ht]
\epsscale{1.25}
 \plotone{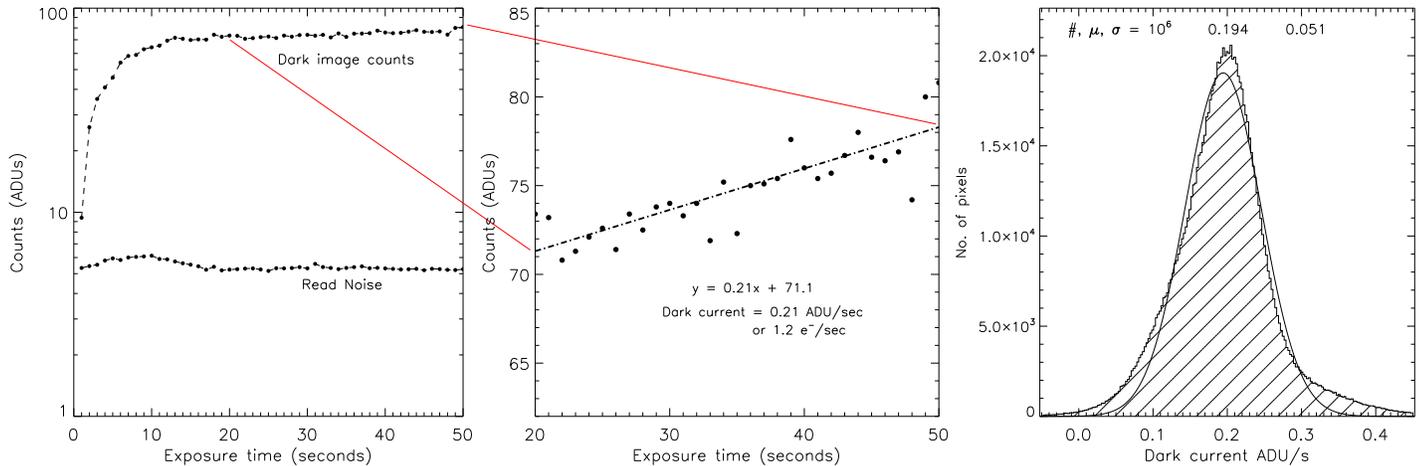}
 \caption{The \textit{left panel} shows log plot of measured dark count value against exposure time for a single pixel. Also shown is the plot of readout noise measured from the BIAS images associated to each dark image. The mean readout noise is obtained to be $5.3\,\mathrm{ADUs}$ or $30\,\mathrm{e^-/pixel}$. The \textit{center panel} shows the section of dark count value, to which a first-order polynomial is fitted for estimating the dark current. The \textit{last panel} shows the histogram analysis for dark current values of all the pixels in the detector. The mean time-dependent dark current is obtained to be $0.2\,\mathrm{ADUs/sec}$ or $1.2\,\mathrm{e^-/sec}$. }
\label{fig3}
\end{figure*}

The left panel in Figure~\ref{fig2} shows the gain measurement plot with variance against signal for one particular box region of 32 $\times$ 32 pixels. Such gain measurements were repeated throughout the field using the same box size covering all the pixels in the detector array. A total of 1024 gain measurements were obtained for all the pixels covered by placing the boxes at different positions. The values were examined by histogram distribution, to which a Gaussian was fitted to obtain the mean gain value of CANICA. The peak of the Gaussian fit and the histogram maximum gave the mean gain value as $5.8\,\mathrm{e^-/ADU}\pm0.8$. The right panel in Figure~\ref{fig2} shows the histogram distribution with the Gaussian fit\footnote{In each plot of Gaussian fit, \# indicates the total number of measurements in the distribution, $\mu$ indicates the peak value of the Gaussian fit, and $\sigma$ indicates the standard deviation of the Gaussian fit.}. The large dispersion in the gain measurements were due to the inclusion of regions which contained effects from vignetting, bad pixels, and other cosmetic effects.

\subsection{Dark current}\label{dcurrent}

The dark current arising in the detector is mainly due to thermal charge generation-recombination and charge diffusion in the semiconductor layers \citep[see \S~8.2]{mclean08}. The dark current is temperature dependent and decays rapidly when cooled to very low temperatures. The detector lab values estimated the dark current to be $<0.1\,\mathrm{e^-/sec}$ for an operating temperature of $78\,\mathrm{K}$. CANICA uses a liquid nitrogen cooling system that maintains the entire cryostat around $77\,\mathrm{K}$.

To estimate the dark current in CANICA, we used images obtained by exposing the camera to a non-illuminated condition by choosing a dark slide in the filter position. Multiple sets of 10 dark images were obtained for exposure times ranging from $1\,\mathrm{s}$ to $50\,\mathrm{s}$ in $1\,\mathrm{s}$ intervals. The RAW and the BIAS images of the darks were saved along with the CDS images. The 10 CDS dark images for a single exposure were averaged to obtain the mean dark image. This was repeated for all exposure times to obtain multiple mean dark images. 

The dark count value for each pixel in the mean dark image were analyzed with respect to their exposure times. The left panel in Figure~\ref{fig3} shows a plot of dark count against exposure time for a single pixel. The dark count is seen to increase rapidly up to $70\,\mathrm{ADUs}$ in $\sim15\,\mathrm{s}$, after which it increases linearly at a slow rate. This latter slow increase is expected to be due to dark current. However, the reason for the rapid increase during first $15\,\mathrm{s}$ is not fully understood, and is believed to be due to reset anomaly. 

To obtain the time-dependent dark current, the dark count value for exposure times ranging from $20\,\mathrm{s}$ to $50\,\mathrm{s}$ was fitted by a first-order polynomial. The slope of the fit represented the dark current value for that particular pixel. This was repeated for all the $1024\times1024$ pixels obtaining large sample of dark current values. The dark current values were then examined by a histogram and a Gaussian was fit to the distribution. The peak of the Gaussian fit gave the mean dark current value. The center panel in Figure~\ref{fig3} shows the dark count value for a single pixel with the first-order fit. The last panel in Figure~\ref{fig3} shows the histogram distribution of dark current values for all the pixels in the detector. The mean time-dependent dark current value from the histogram is obtained as $0.2\,\mathrm{ADUs/sec}$ or $1.2\,\mathrm{e^-/sec}$. The low levels of dark current do not affect the image quality, but images with exposure time more than $30\,\mathrm{s}$ should be carefully reduced as the dark current will add up to be more than the readout noise (see Section~\ref{readn}). Extrapolating the dark current value and using it for different exposures is not recommended. Observers need to obtain the darks corresponding to their object exposures each night for accurate image reduction.

\subsection{Readout noise} \label{readn}

Readout noise is the total amount of noise generated by the camera electronics (capacitors, amplifiers, ADC, etc.) when the detected charge in $e^{-}$ is transferred and measured as digital units (ADUs) \citep{jane01}. The readout noise determines the ultimate performance of the camera electronics and cannot be completely eliminated, but can be minimized. Low readout noise have been achieved in infrared arrays using different readout techniques such as Fowler sampling \citep{fowl90}. With operating temperature of $77\,\mathrm{K}$ and using the CDS readout technique, we expected CANICA readout noise to be close to the lab value of $10\,\mathrm{e^-/pixel}$.

The readout noise in an image can be obtained as the standard deviation of all the pixel values when other noise contributions are removed. The measurement of readout noise for CANICA was carried out using the BIAS images obtained from the dark exposures (see Section~\ref{dcurrent}), by a series of steps as follows: \\
\indent 1) The 10 BIAS images obtained at each exposure time were averaged to obtain the mean BIAS image (the BIAS images do not depend on exposure time since they are the first ``readout'' image after the reset signal. We are using all the BIAS images for a statistical analysis).\\
\indent 2) The mean BIAS image was used to compute the difference with each of the 10 individual BIAS images in the set. This produced 10 differenced images for a particular exposure time (differencing two BIAS images removes the $Reset$ or $``kTC"$ noise \citep[see \S~6.3]{jane01} and leaves only readout noise contribution). \\
\indent 3) The differenced images were divided by $\sqrt{1.1}$ (see Appendix~\ref{appB}) to account for noise increase during subtraction.\\
\indent 4) Next, each differenced image was examined by a histogram with a Gaussian fit. The standard deviation of all the pixels from the Gaussian fit was taken as the readout noise. This produced 10 readout noise values. \\
\indent 5) The 10 readout noise in each set were median filtered to obtain the final readout noise value for a particular exposure time.\\
\indent 6) Steps 1 to 5 were repeated for all the sets of BIAS images to give a series of final readout noise values at different exposure times.

The left panel in Figure~\ref{fig3} shows the final readout noise against different exposure times. The final readout noise were average combined to obtain the mean readout noise of CANICA, which was estimated as $5.3\,\mathrm{ADUs}$ or $30\,\mathrm{e^-/pixel}$. The HAWAII array in CANICA is made up of four quadrants with different readout electronics for each quadrant. Analyzing readout noise by individual quadrants, we see that quadrants Q\Rnum{2}, Q\Rnum{3} and Q\Rnum{4} have values similar to the mean readout noise of $30\,\mathrm{e^-/pixel}$. Whereas the quadrant Q\Rnum{1} shows a higher readout noise of $65\,\mathrm{e^-/pixel}$. 

\subsection{Linearity corrections} \label{lin}

\begin{figure*}[!ht]
\epsscale{1.15}
 \plotone{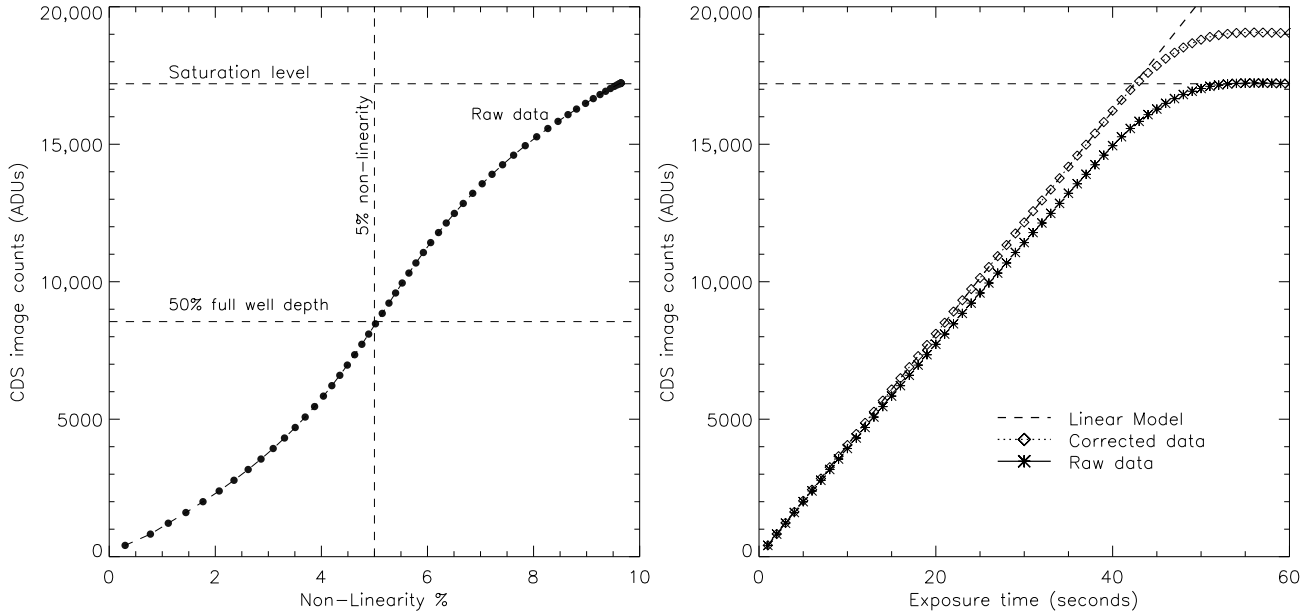}
 \caption{The \textit{left panel} shows the non-linear behavior of the detector obtained from the observed counts in raw CDS images. The detector is 5\% non-linear for 50\% of the full-well-depth. The \textit{right panel} shows linearity corrected response for a single pixel in a CDS image. The uncorrected count values are shown with the star symbol, the linearity corrected count values with the diamond symbol and the linear model is represented by a dashed line. }
   \label{fig4}
\end{figure*}

The pixels in the HAWAII detector array have an inherent non-linear response to light due to changes in detector capacitance which are caused by the change in bias voltage at different levels of pixel integration \citep{jane01}. Hence, the measured counts of a bright source can be significantly less than the true counts and this affects photometric quality of the data. Accurate methods for non-linearity correction have been advocated by \citet{vacca04} and \citet{clemens07} for NIR arrays. Based on these studies, we developed a simple technique that is faster and useful to correct the non-linearity to better than 1\% levels for a CDS image. 

The CDS image is formed by the difference of RAW and BIAS images. \citet{clemens07} emphasized that some pixels in the BIAS image can already be into the non-linear portion of their response when there is a bright source. However, based on our analysis of various bright sources within the saturation limit of the detector, we see that the illumination caused by bright sources in the BIAS image is less than 10\% of the full-well-depth. This value is very low and would fall within 1\% of the non-linear regime. Hence, in our linearity correction method we use only the CDS images for correction (i.e.\ we ignore non-linear effects in the BIAS images). The left panel in Figure~\ref{fig4} shows the non-linear response of CANICA. It is seen that the detector has 10\% non-linearity near the full-well-depth. This decreases to 5\% for half the well depth and 1\% for one-tenth of the well-depth.

The linearity correction method involves characterizing the illumination response of each pixel in the image. This is carried out by collecting various number of dome flat images at different exposure levels for a fixed illumination and is described as follows:\\
\indent 1) We obtained multiple sets of 10 dome flats in $H$-band for exposure times ranging from $1\,\mathrm{s}$ to $60\,\mathrm{s}$ in increments of $1\,\mathrm{s}$. \\
\indent 2) The 10 flats at each exposure time were averaged. This produced 60 mean flat images corresponding to each exposure time. The mean flat images were examined per pixel with a plot of their observed counts against exposure time. After a certain time, the count level reaches a saturation value after which linearity correction cannot be applied (see Figure~\ref{fig4}).\\
\indent 3) For the plot of observed counts against exposure time for each pixel, a fourth-order polynomial was fitted up to 90\% of the saturation value to obtain the non-linear behavior of the pixel. The fourth-order polynomial can be expressed as
\begin{equation}
Ct = c_{0} + c_{1}T_{e} + c_{2}T_{e}^2 + c_{3}T_{e}^3 + c_{4}T_{e}^4 
\end{equation}
where $T_{e}$ is the exposure time, $Ct$ is the observed counts and $c_{0},c_{1},c_{2}..$ are the coefficients of the fit.\\
\indent 4) The first two coefficients of the fourth-order fit gives the best representation of the linear response of the pixel. With this we constructed the linear model of the pixel as
\begin{equation}
linear_{model} = c_{0} + c_{1} T_{e}
\end{equation}
\indent 5) Next, for the plot of linear model against observed counts another fourth-order polynomial was fitted: 
\begin{equation}
linear_{model} = c'_{0} + c'_{1}Ct + c'_{2}Ct^2 + c'_{3}Ct^3 + c'_{4}Ct^4 
\end{equation}
The new coefficients of this fit gives the correction terms needed for linearity correction. The coefficients for all the pixels in the detector were then saved into an image array and named as $ic'_{0}, ic'_{1}, ic'_{2}, ic'_{3}$ and $ic'_{4}$. The saturated pixels and bad pixels are ignored and not taken into account during the correction.\\
\indent 6) Correction for non-linearity was then applied by using the new coefficients to a CDS image as follows:
\begin{equation}
CDS_{corr} = ic'_{0} + ic'_{1}CDS + ic'_{2}CDS^2 + ic'_{3}CDS^3 + ic'_{4}CDS^4 
\end{equation}

The linearity correction results for multiple CDS images obtained from various dome flats at different exposure times are shown in right panel of Figure~\ref{fig4}. The corrected and uncorrected data for one particular pixel are plotted against the exposure time. The linearity corrected data are shown with the diamond symbol and uncorrected data are shown with the star symbol, with the linear model represented by a dashed line. 

\subsection{Crosstalk}

The HAWAII array displays some anomalous behavior for images with high count levels. One commonly seen problem is electrical crosstalk. The effect is that spurious negative ghost images of a bright source appear in different readout channels \citep[see \S~7.2]{jane01}. The problem arises when different output amplifiers are read in parallel while drawing power from a common supply line. The signal from one channel couples into other channels due to voltage leakage causing change in the drain resistance of the amplifier. The crosstalk effect in CANICA is observed in the four quadrants of a CDS image as dark patches and are dubbed ``holes''. The holes have a bleeding effect during readout, hence there is a trail left along the read direction. \citet{finger08} showed that the effect of crosstalk diminishes with slower clock speed. For CANICA, we evaluated the effect of crosstalk at different full-well-depths in multiple images. It was found that the amount of negative signal in different quadrants is around 0.8\% of the bright source. We incorporated a custom program in the reduction pipeline for correcting crosstalk effects. The program is able to remove the holes by adding to the pixels corresponding to other quadrants with 0.8\% of the peak value of the source. This considerably improves the image quality and the effects are corrected below 1\% levels. 

\section{Observing scheme and image reduction} \label{obsch}

CANICA observations include various effects that depend on time caused by both local and global effects, introduced by the atmosphere and by the instrument. At NIR wavelengths there is strong sky contribution from telluric lines and thermal emission. Instrumental effects include variations of thermal emission from the local structures due to variation in the ambient temperature. The former variation is expected to be uniform over the FOV, whereas the latter could be pixel dependent. The combined effect of both of these variations can be determined by obtaining ``sky'' image for each set of observation. 

The key aspect of an observing scheme is to create a telescope dither sequence that facilitate preparing a sky image, which can be subtracted during reduction process. The CANICA scheme uses a dither methodology to obtain multiple images for a given observing field. Typically, 15 images are obtained for a source, to result in a combined high S/N image.
The observing scheme is separated into two types depending on the size of the source:\\
\indent 1) For extended sources, the sequence of observation consists of images with dithering pattern of alternating source and off-field images. The off-field image is obtained by dithering outside the source field in different cardinal directions (typically in north-south direction). \\
\indent 2) For point sources, the sequence of observation consists of images obtained with dithering distributed in a non-repetitive random pattern within a diameter of 20 to 30 arcseconds around the targeted offset.

Typical exposure times range from $1\,\mathrm{s}$ to $120\,\mathrm{s}$ for bright to faint sources in the $H$-band. The software control in the host computer coordinates the observation sequence and the user passes a Java script with instructions for image acquisition and dithering to the telescope control system.

Image reduction for CANICA is carried out using standard NIR reduction technique, which involves dark subtraction, flat fielding, and sky removal as per the equation below:
\begin{equation}
\rm Image = \frac{CDS_{corr} - Dark}{nFlat} - Sky
\end{equation}

Before the image reduction process, the CDS images in each observation are linearity corrected to obtain $\rm CDS_{corr}$. The first step in the image reduction is dark count removal. To obtain a good estimate of the dark count, a sequence of 10 darks are obtained for the same object exposure at the end of the night. These are then averaged to get the mean dark image which is used for dark subtraction. Next, flat fielding is carried out by dividing with the normalized flat, to correct the pixel-to-pixel variations and illumination profile. The flats are obtained from dome flat screen with lights ON and OFF technique, which are differenced. The differenced flat is normalized by using the mean value of a large sample of pixels spread across the central $2 \times 2\,\mathrm{arcmin^2}$. After flat fielding, the sky image is estimated by stacking and taking the median of the dithered images. The resulting sky image is then used for subtracting the sky contributions from the flat-fielded image. The sky subtracted images at different dither positions are then aligned and averaged to obtain the final image for a given observing field. A custom pipeline\footnote{The reduction pipeline along with several other scripts, developed by one of us (Y.D.M.) is downloadable as a package at \url{http://www.inaoep.mx/~ydm/inaoe_iraf.html}} incorporating this reduction scheme is developed locally under the IRAF\footnote{Image Reduction and Analysis Facility (IRAF) is distributed by the National Optical Astronomy Observatory, which is operated by the Association of Universities for Research in Astronomy (AURA) under a cooperative agreement with the National Science Foundation: \url{http://iraf.noao.edu/}} environment.

\subsection{Photometry and calibration} \label{photcal}

Photometric calibration for CANICA images is performed robustly in each of the $J$, $H$, and $K$\footnote{The CANICA broadband filter $K^{'}$ is calibrated with the 2MASS filter $K_{s}$. Hence, all values are specified according to the 2MASS $K_{s}$ wavelength. Figures, tables, and results are represented as ``$K$'' filter instead of $K^{'}$ or $K_{s}$ to avoid confusion, unless specified.} broadbands to extract accurate astrometry and magnitudes of the sources. The 2MASS \citep{skrut06} data set available publicly is used for calibration purposes. 

The final reduced images are first astrometrically corrected before performing photometry. Astrometry values available in the CANICA image headers have source position offset by a few arcseconds to a few arcminutes when compared with their true coordinates. The images were also found to have rotation offsets of a few degrees and to possess slight geometric distortions. Hence, the astrometry correction involved careful analysis using a customized program. The first step in the analysis involves coarse corrections by copying the 2MASS coordinates of a reference star into the image header. The second step involves obtaining solutions to rectify the image rotation and geometric distortions. This is performed with the help of tasks in IRAF \textit{imcoords} package. A minimum of six sources within the field are chosen obtaining their centroids and 2MASS coordinates. Next, the centroids and coordinates are matched to compute the final plate solutions. The header information in each image is then transformed appropriately to yield images corrected for astrometry in the equatorial coordinate system.

\begin{figure}[!ht]
\epsscale{1.2}
\plotone{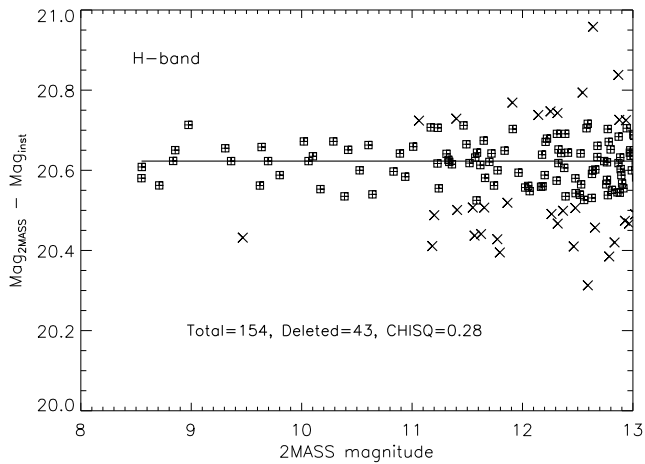}
 \caption{Plot of zeropoint measurements against magnitude for a single observing field in $H$-band. The individual stellar zeropoints ($\rm ZP_{star}$), obtained as the difference of 2MASS and instrumental magnitude are represented by box symbols. The horizontal line in the plot represents the zeropoint of the field ($\rm ZP_{field}$). The data points excluded in the measurements are shown by cross symbols.}
   \label{fig5}
\end{figure}

Once images are astrometrically corrected, aperture photometry is performed in IRAF using the \textit{digiphot} package on all the point sources to obtain flux, magnitudes, and their errors. An aperture radius of 10 pixels (2*FWHM, see~\ref{PSF}) is considered best for photometry based on magnitude growth analysis \citep{howell89} with respect to different aperture radius. The photometric annulus is chosen to be around 25 pixels with the dannulus of 10 pixels. After photometric analysis, the values of source centroids with their corresponding magnitudes are saved in a file. The instrumental magnitude obtained during photometry is calculated as
\begin{equation}
\rm Mag_{inst} = -2.5log(counts/sec)
\end{equation}
The individual stellar zeropoint ($\rm ZP_{star}$) in the observed field is determined by comparing the instrumental magnitudes to their equivalent 2MASS magnitudes as
\begin{equation}
\rm ZP_{star} = Mag_{2MASS}-Mag_{inst}
\end{equation}
The zeropoint for the observed field ($\rm ZP_{field}$) is then obtained by averaging all the stellar zeropoints of ``good stars,'' by visually eliminating outliers. The process is summarized in Figure~\ref{fig5}, where the horizontal line shows the zeropoint for the field and the eliminated stars are shown by cross symbols.

\section{CANICA Performance}

CANICA was designed as an imaging camera with a single operating mode. The only change during observations is to the filter configuration. Hence, all of the optical setup with the detector array was fixed. This helped to evaluate the camera performance using data from multiple observations at different nights. We used data that were obtained after re-aluminization of the telescope mirror during 2016 September, totaling 13 nights of observations. In the following sections, we present various imaging performance parameters of CANICA. 

\subsection{Seeing and Point Spread Function}\label{PSF}

The typical atmospheric seeing at the observatory location\footnote{\url{http://astro.inaoep.mx/observatorios/cananea/elobservatorio/condiciones.php}} is $\sim0.{\arcsec}9$. The seeing value changes between each of the primary broadbands, as $\lambda^{-0.2}$ \citep{fried66}. The response of an imaging system to a point source is described by the PSF and is measured as the full width at half maximum (FWHM). The major contribution to the PSF in a ground-based observation is atmospheric seeing. However, instrumental effects can contribute to additional changes in the PSF. The PSF for a linear imaging system should be constant and not depend on the magnitude of the source (as long as it is not saturated). Hence, if the FWHM measurements varied both in time and across the FOV, then this behavior is believed to be due to two main reasons: \\
\indent 1) Changes in the atmospheric seeing at different time of observations.\\
\indent 2) Result of aberrations in the image quality across the detector's FOV. 

\begin{figure}[!ht]
\epsscale{1.2}
\plotone{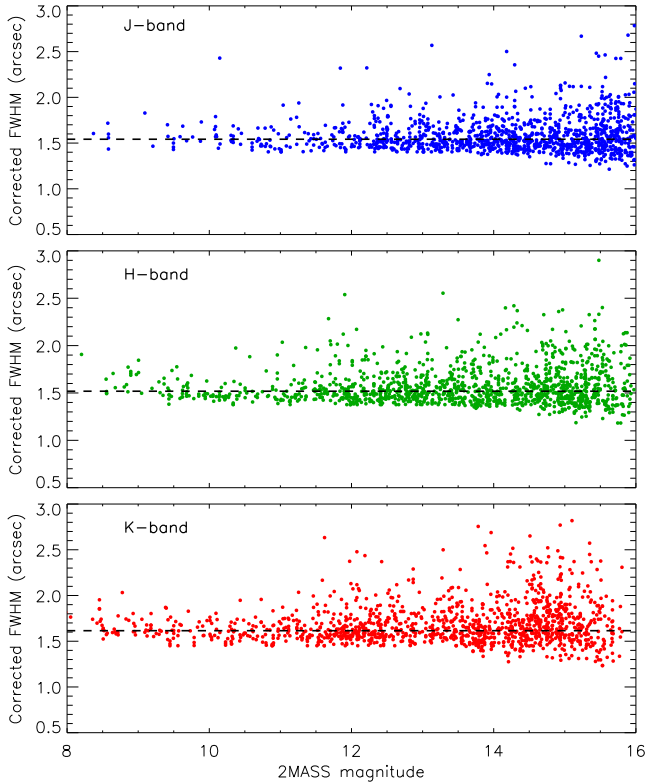} 
 \caption{Plot of corrected FWHM measurements against magnitude in $J$, $H$, and $K$ broadbands. The FWHM values are filtered for sources located within the detector's central FOV of $4 \times 4\,\mathrm{arcmin^2}$. The dashed line in each plot represents the median FWHM value.}
   \label{fig6}
\end{figure}

Considering the above statements and for a given constant seeing condition, one can measure the variations in PSF values due to aberrations in the camera.

To obtain PSF of CANICA, we analyzed 43 observations, spanning 13 nights, taken in all the primary broadbands. The data set consisted of images of photometric standards, open clusters, and regions from the nearby galactic zone. The point source selection was carried out using IRAF \textit{daofind} task, with a detection threshold of $5\sigma$. In total, 3142 sources were obtained from all the observations having 2MASS matches. The PSF for each source was fitted with a Gaussian profile and its FWHM was measured. As the seeing changed across different nights, the FWHM for each observing field were normalized and scaled to provide corrected constant seeing. Normalization and scaling is performed as follows: 

\begin{figure}[!ht]
\begin{center}
\epsscale{0.98}
\plotone{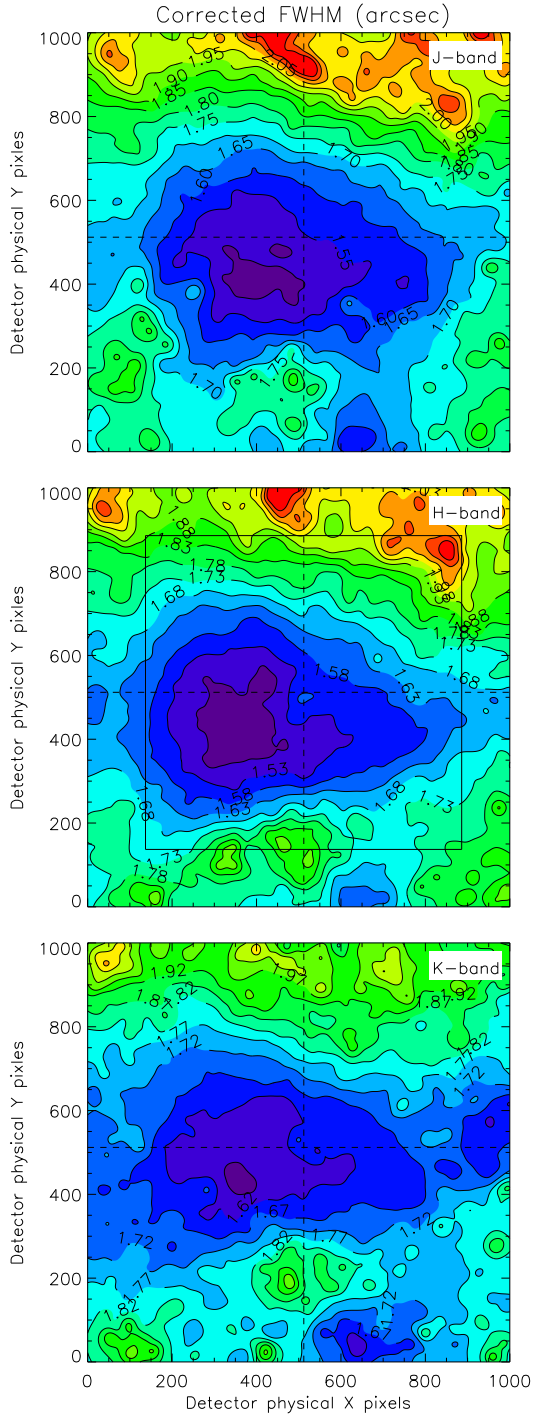}
 \caption{Variations in PSF (corrected FWHM, see Section~\ref{PSF}) across the detector's full FOV ($5.5\times5.5\,\mathrm{arcmin^2}$ or 1024 $\times$ 1024 pixels) in $J$, $H$, and $K$ broadbands. The contours start at 1.45, 1.43, 1.50 for $J$, $H$, $K$ and are stepped in increasing order of 5\% values up to 15 contour levels. The central $4\times4\,\mathrm{arcmin^2}$ region is shown by a black box in $H$-band panel.}
   \label{fig7}
   \end{center}
\end{figure}

\indent 1) For a single observing field, the minimum value of all the FWHM measurements for magnitudes brighter than $13\,\mathrm{mag}$ across the central FOV of $4 \times 4\,\mathrm{arcmin^2}$ was obtained (This minimum value represented the true PSF for a given observing field, as bright sources falling on the central FOV would have least effect due to aberrations). \\
\indent 2) The minimum FWHM value was then used to normalize (division) the FWHM values for all the sources in a given observing field.\\
\indent 3) Steps~1 and 2 were repeated for all observing fields at different nights to obtain a large sample of normalized FWHM values.\\
\indent 4) All the normalized FWHM values were then scaled to the mean FWHM value (obtained by taking average of values from Step~1 for all nights) by multiplying with constants of $1.{\arcsec}33$, $1.{\arcsec}30$, and $1.{\arcsec}37$ for $J$, $H$, and $K$ broadbands. These values now formed the corrected FWHM measurements for the 3142 sources spread across different locations on the detector.

The corrected FWHM measurements were compared with the source magnitudes obtained from 2MASS data. Only sources within the central FOV of $4 \times 4\,\mathrm{arcmin^2}$ were chosen. Figure~\ref{fig6} shows the plot of FWHM values against magnitude for all the primary broadbands. The dashed line in the figure represents median corrected FWHM value. The median FWHM value is obtained to be $\sim1.5\,\mathrm{\arcsec}$ or 5~pixels in all the filters. It is seen that there is increasing scatter in FWHM variations for fainter sources. The asymmetric scatter at fainter magnitude could be due to increasing contamination of background extended sources (such as galaxies) in our sample.

To obtain the PSF variations across the FOV, the corrected FWHM values were examined by their position in the detector. Only sources brighter than $14\,\mathrm{mag}$ were chosen for reliability. Figure~\ref{fig7} displays the contour map of corrected FWHM values for all the primary broadbands. The different color and contour levels depict the variation in FWHM values across the full FOV. It is seen that there is a radial gradient of increasing FWHM values from the center toward the edges, observed in all the three filters. The level of FWHM changes are around 10\% (or $0.1\,\mathrm{arcsec}$) within the central $4\times4\,\mathrm{arcmin^2}$ region. The variations in FWHM values can be attributed as a result of coma and vignetting from the camera optics. The effect of coma at the edges of the detector is larger than the PSF variations and is not completely seen in Figure~\ref{fig7}. Analysis of two-dimensional PSF profiles of sample stars near the edges give a FWHM difference in X and Y direction of around $0.5\,\mathrm{arcsec}$, whereas stars near the central FOV have a difference less than $0.2\,\mathrm{arcsec}$. Thus, the central $4 \times 4\,\mathrm{arcmin^2}$ is the best in terms of imaging quality and is recommended as the FOV to use for observations.

\begin{figure}[!ht]
\epsscale{1.2}
\plotone{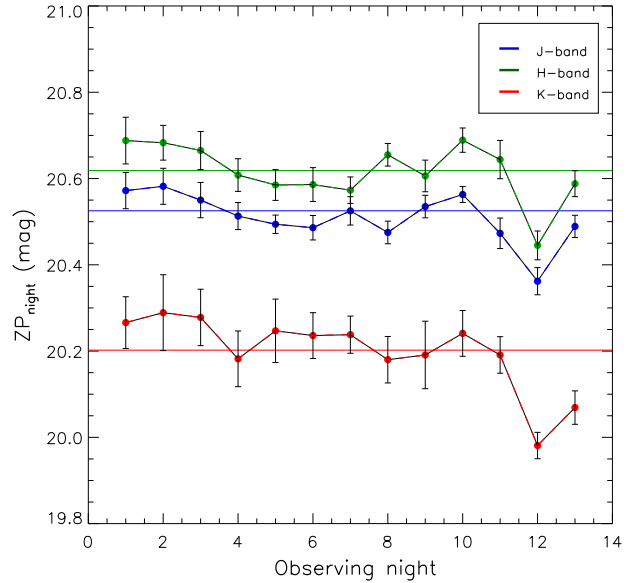} 
 \caption{Plot of nightly zeropoint ($\rm ZP_{night}$) values against different observing nights in $J$, $H$, and $K$ broadbands. The $\rm ZP_{night}$ values are obtained from the average of $\rm ZP_{field}$ values in each night. The straight lines  corresponding to each filter represents the average zeropoint value of CANICA ($\rm ZP_{CANICA}$).}
   \label{fig8}
\end{figure}

Another aspect observed in the corrected FWHM map is the location of the minimum contour level. The minima signifies the position of the optical axis on the FOV. The minimum FWHM values tend to be distributed near the central FOV but are more biased toward the first quadrant of the detector. This indicates that the optical axis is slightly off-centered on the detector's FOV. Overall, we find that the PSF of CANICA changes across the full FOV by $\sim20\%$, with $\sim10\%$ changes in the central $4\times4\,\mathrm{arcmin^2}$ FOV. 

\begin{figure*}[!ht]
\epsscale{1.2}
\plotone{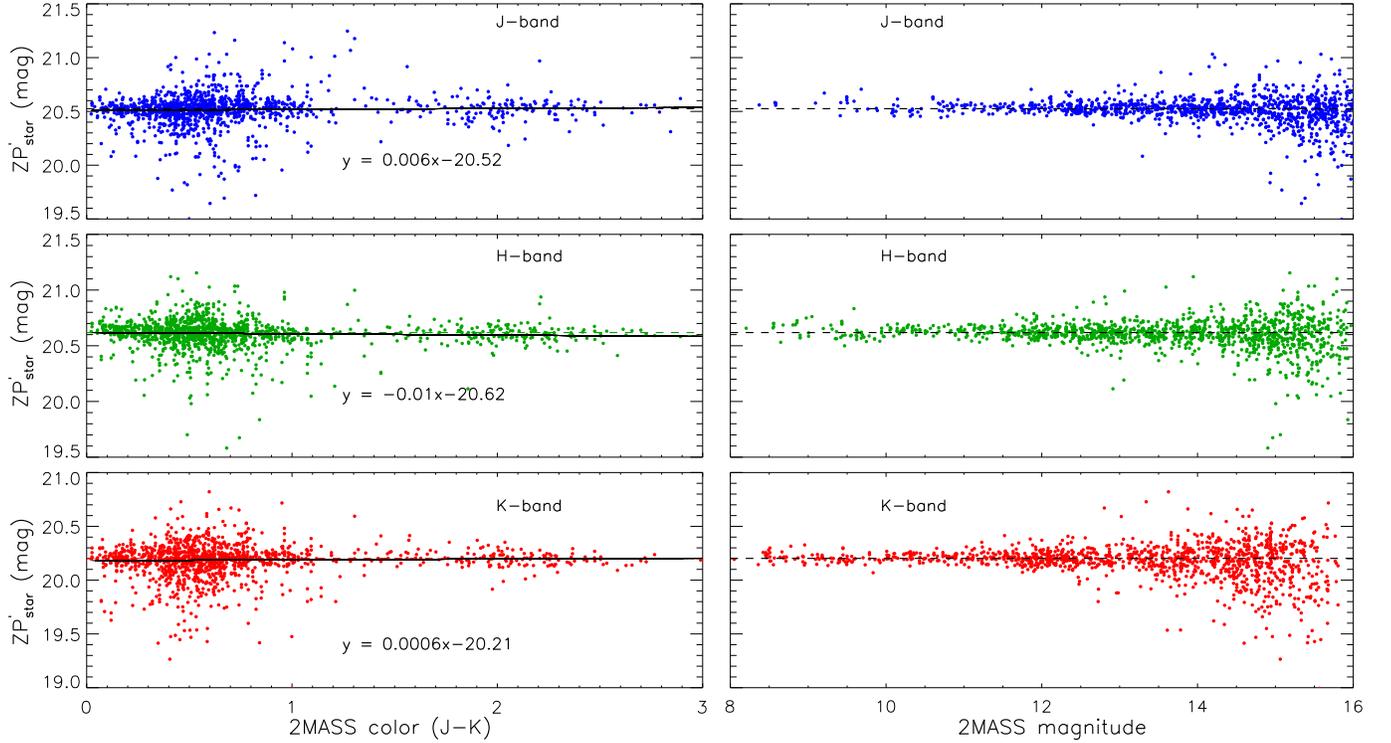}
 \caption{The three \textit{left panels} and the three \textit{right panels} show plots of corrected stellar zeropoint ($\rm ZP'_{star}$) values against color and magnitude in $J$, $H$, and $K$ broadbands. In the \textit{left panels}, the $\rm ZP'_{star}$ values are fitted with a first-order polynomial. The slope of the fit is negligible in all the three bands indicating the filters are standardized in their respective bandwidths and central wavelengths. The \textit{right panels} show $\rm ZP'_{star}$ dispersions with magnitude. The dashed line represents the average zeropoint value of CANICA ($\rm ZP_{CANICA}$).}
   \label{fig9}
\end{figure*}

\subsection{Zeropoint variations}\label{ZPvar}

We carried out photometric analysis of point source fields to obtain zeropoint values for 43 observations, spanning 13 nights, in all the primary broadbands (the observations belonged to the same data set used for studying PSF variations in Section~\ref{PSF}, which totaled 3142 2MASS matched sources). For each observing field, its $\rm ZP_{field}$ was obtained as described in Section~\ref{photcal}, using $\rm ZP_{star}$ values for sources brighter than $13\,\mathrm{mag}$ across the central $4\times4\,\mathrm{arcmin^2}$ region. The $\rm ZP_{field}$ for each night were averaged to obtain the zeropoint value for that night ($\rm ZP_{night}$). This was repeated for all nights to obtain zeropoint variations over time. Figure~\ref{fig8} displays zeropoint values for 13 nights of observations in all the primary broadbands. The RMS over each $\rm ZP_{night}$ is around $0.05\,\mathrm{mag}$. Night-to-night variations of the zeropoints was found to be within $0.1\,\mathrm{mag}$, except for the last two nights. Average combining all the $\rm ZP_{night}$ values gave us the average zeropoint of CANICA ($\rm ZP_{CANICA}$) in the three broadbands as $J = 20.52\pm0.05$, $H = 20.63\pm0.06$, and $K = 20.23\pm0.08$.

\begin{figure}[!ht]
\epsscale{1.02}
\plotone{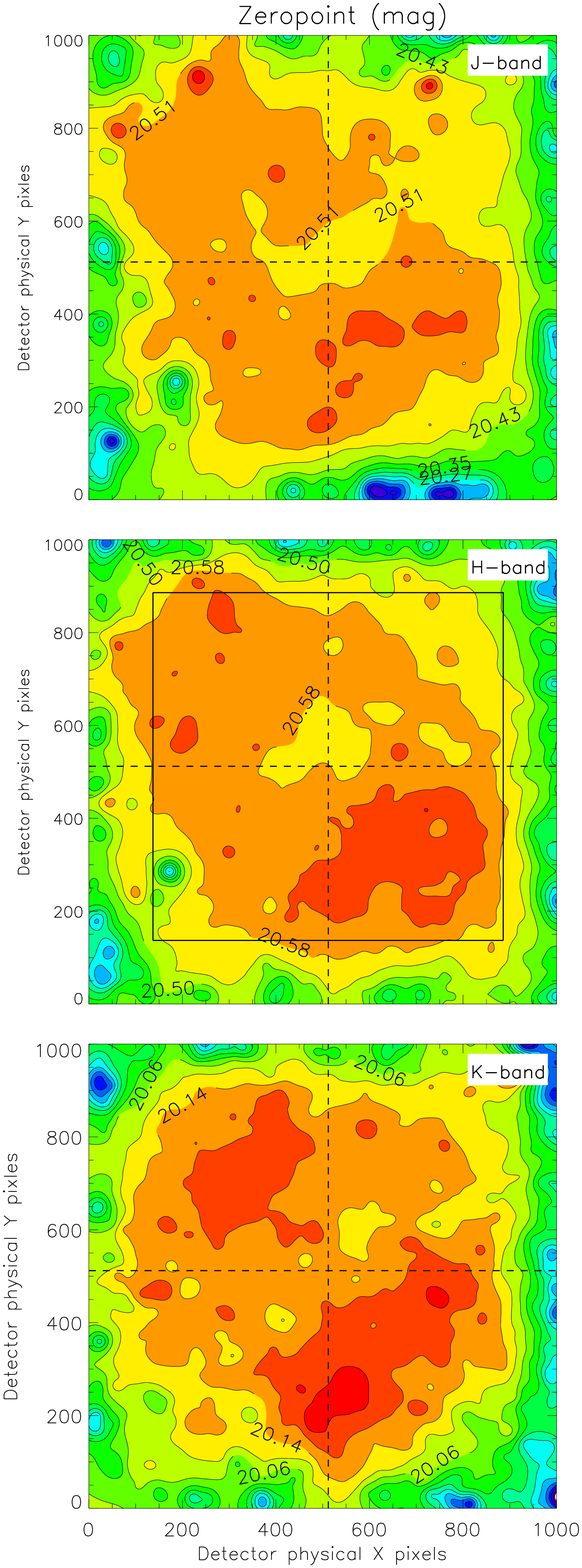} 
 \caption{Variations in zeropoint values across the detector's full FOV ($5.5\times5.5\,\mathrm{arcmin^2}$ or 1024 $\times$ 1024 pixels) in $J$, $H$, and $K$ broadbands. The contours start at 20.03, 20.10, 19.65 for $J$, $H$, $K$ and are stepped in increasing order of 4\% values up to 15 contour levels. The central $4\times4\,\mathrm{arcmin^2}$ region is shown by black box in $H$-band panel.}
   \label{fig10}
\end{figure} 

To study variations in zeropoint values across the detector's FOV, we combined all the 3142 individual zeropoint ($\rm ZP_{star}$) values of multiple observations in all the primary broadbands. Because the zeropoints changed for each observing field, they had to be corrected before combining all the measurements. The correction was carried out using the following equation:
\begin{equation}
\rm ZP'_{star} = (ZP_{star} - ZP_{field}) + ZP_{CANICA}
\end{equation}
where $\rm ZP'_{star}$ represents individual stellar zeropoint corrected for field-to-field and night-to-night variations.

Comparative study of $\rm ZP'_{star}$ with respect to their color and magnitude was carried out in all the primary broadbands. Only sources within the central FOV of $4 \times 4\,\mathrm{arcmin^2}$ were chosen in this study. The left panels in Figure~\ref{fig9} show the plot of $\rm ZP'_{star}$ values against color ($J-K$). A linear fit to the $\rm ZP'_{star}$ vs $J-K$ helps us to evaluate a possible color coefficient of the zeropoint. The values of slope calculated in $J$, $H$, and $K$ are 0.006, 0.01, and 0.0006. These values are relatively small indicating the CANICA filters are matched well to the 2MASS filters in their respective bandwidths and central wavelengths. The right panels in Figure~\ref{fig9} show the plot of $\rm ZP'_{star}$ values against 2MASS magnitude. The zeropoints for sources brighter than $13\,\mathrm{mag}$ have dispersions less than $0.05\,\mathrm{mag}$. For fainter sources, the dispersion in zeropoint increases reaching values of $\sim0.2\,\mathrm{mag}$ in $J$, $H$, and $K$ bands.

The variation in zeropoint values across the detector's FOV were obtained by examining the source positions with respect to their $\rm ZP'_{star}$ values. Only sources brighter than $14\,\mathrm{mag}$ were chosen in this study. Figure~\ref{fig10} displays the contour map of $\rm ZP'_{star}$ values for all the primary broadbands. The changes in zeropoint values are consistent, with variations less than $0.1\,\mathrm{mag}$ across the central FOV of $4 \times 4\,\mathrm{arcmin^2}$. At the edges of the detector there is a strong roll off in the zeropoint values. The zeropoints variations at the edges are due to flux loss for coma dominated sources, since a fixed photometric aperture is used (which is based on mean PSF). Unlike the variations in FWHM values, there is no radial change in zeropoint profiles from the center, indicating that there is no effect due to aberrations or PSF changes. Hence, the zeropoint values have no large dependency based on the source positions. The second quadrant Q\Rnum{2} is seen to have slightly lower zeropoint of around $0.05\,\mathrm{mag}$ when compared with other quadrants. This may indicate the quantum efficiency of quadrant Q\Rnum{2} is lower than the other quadrants. In Section~\ref{readn}, it was estimated that quadrant Q\Rnum{1} has higher readout noise. The zeropoint values in Q\Rnum{1} are consistent showing photometry is not affected by readout noise. We also noted that the optical axis on the detector is slightly off-center, but it is seen that the zeropoints are constant near the central FOV. Overall, we find that the zeropoints have very low variations throughout the central FOV of $4 \times 4\,\mathrm{arcmin^2}$. This allows accurate photometry with CANICA.

\subsection{Throughput}

Throughput gives the transmittance of photons through the earth's atmosphere, telescope, and the camera system. Throughput can be estimated by combining the transmission efficiencies of each of the systems involved. A better way to obtain throughput is from observations. This involves measuring the ratio of the incident number of photons per second outside the earth's atmosphere to the detected number of photons per second by the camera, for a standard star on a given telescope aperture at a particular wavelength. 

To obtain CANICA's throughput, we performed both the theoretical and observational analyses. The theoretical estimation includes the values of atmospheric transmission, the reflectivity of the primary and secondary mirrors of the telescope, the transmission of the camera window, the transmission of the camera optics, filter transmission and the detector quantum efficiency. These values were derived from lab manuals and the literature and are summarized in Table \ref{tbl-2}. The total throughput is then calculated by multiplying all the individual system transmissions. We find the theoretical estimates of total throughput in all the primary broadbands to be $J$ = 15.0\%, $H$ = 16.1\%, and $K$ = 19.8\%.

\begin{deluxetable*}{lccc}
\tablecolumns{4}
\caption{CANICA throughput.}\label{tbl-2}
\tablehead{
\colhead{Item}  & \colhead{$J$-band} & \colhead{$H$-band} & \colhead{$K$-band}\\ 
}
\startdata
Central wavelength $\lambda_{c}$ ($\mu$m)  	& 1.246				& 1.633			& 2.119 \\
Bandwidth $\Delta\lambda$ ($\mu$m)  			& 0.163 				& 0.296			& 0.351 \\
Absolute flux $F_{\lambda}(0)$ (Wcm$^{-2}$ $\micron^{-1}$)		& 2.94 $\times$ 10$^{-13}$ 		& 1.14 $\times$ 10$^{-13}$ 		& 3.89 $\times$ 10$^{-14}$  \\
\hline
Atmospheric transmission (\%)  				& 90 				& 93			& 91   \\
Telescope reflectivity (\%)  				& 81	 				& 83			& 85   \\
Filter transmission (\%)  					& 84 				& 81			& 88   \\
Camera optics transmission (\%)				& 43	 				& 46			& 47   \\
Detector QE (\%)		  						& 57 				& 56			& 62   \\
\hline
Total transmission (estimated) (\%)  		& 26.3	 			& 28.7			& 31.9  \\
Total transmission (measured) (\%)  			& 16.63	 			& 20.1			& 23.9 \\
\hline
Total throughput (estimated) (\%)  			& 15.0	 			& 16.1			& 19.8  \\
Total throughput (measured) (\%)  			& 9.48	 			& 11.26			& 14.84 \\
\enddata

\end{deluxetable*}

Measuring throughput values from observations is based on assessing the signal collected by a telescope at a particular wavelength for a source of given apparent magnitude, transmitted by an optical system onto a detector. This requires values of conversion gain ($g$), zeropoint magnitude ($m_{zp}$), filter central wavelength ($\lambda_{c}$), filter bandwidth ($\Delta\lambda$), and absolute flux of a zero-magnitude star $F_{\lambda}(0)$. 

For a signal collected by a telescope at a rate of $1\,\mathrm{ADU/s}$, the equation for zeropoint magnitude can be written as \citep[see Section~9.9]{mclean08}: 

\begin{equation}
m_{zp} = 2.5log(\frac{\tau \eta \lambda_{c} \Delta\lambda A_{tel} F_{\lambda}(0)}{hcg})
\end{equation}

where $\tau$ is the transmission of all systems, $\eta$ is the detector quantum efficiency, $A_{tel}$ is the area of telescope mirror (here $A_{tel}$ = 32300 cm$^2$, for effective primary mirror diameter of $2.05\,\mathrm{m}$ with a central hole of $30\,\mathrm{cm}$ diameter), $h$ is Planck's constant, and $c$ is the speed of light.

The above equation can be rewritten as
\begin{equation}
2.5log({\tau\eta}) = m_{zp} - 2.5log(\frac{\lambda_{c} \Delta\lambda A_{tel} F_{\lambda}(0)}{hcg})
\end{equation}

From this equation, we can calculate the parameter $\tau\eta$, which gives the total throughput of the system. The parameters in the right-hand side of the equation are the known values from previous measurements and the literature. The values of absolute flux of a zero-magnitude star $F_{\lambda}(0)$ were obtained from \citet{hewett06} and is given in Table \ref{tbl-2}. Substituting the values for each parameter, we obtain the total throughput in all the primary broadbands as $J$ = 9.48\%, $H$ = 11.26\%, and $K$ = 14.84\%. When the detector quantum efficiency is excluded, we get the transmission of all systems as $J$ = 16.63\%, $H$ = 20.1\%, and $K$ = 23.9\%. The measured throughput appears lower than the theoretical estimates, mostly due to the difficulty in obtaining accurate atmospheric transmission at the site when there is loss from scattering due to dust and aerosols \citep{carb98,ortiz02}. Overall, the measured values are consistent and indicate the efficiency of the the instrument.

\begin{deluxetable*}{ccccccc}
\tablecolumns{6}
\caption{CANICA photometric performance.}\label{tbl-3}
\tablehead{
\colhead{Broadband}  & \colhead{Limiting Magnitude} & \colhead{Zeropoint} & \colhead{Sky Counts} & \colhead{Sky Magnitude} & \colhead{Sky Saturation} \\ 
\colhead{Filters}  	& \colhead{(for S/N = 10 at $900\,\mathrm{s}$)} & \colhead{Mag} 	& \colhead{(ADUs/sec/pixel)} & \colhead{(mag/sec/arcsec$^2$)}	& \colhead{Exposure Time} \\
& \colhead{Estimated \quad Measured} & & &
}
\startdata
$J$ 			& 18.5 \quad\quad\quad 18.0		& 20.52 		 	& 20 	 & 14.7 		&  	$900\,\mathrm{s}$ \\	
$H$ 			& 17.6 \quad\quad\quad 17.5		& 20.63	 		& 60     & 13.6     &	$300\,\mathrm{s}$ \\		
$K$ 			& 16.0 \quad\quad\quad 16.4	 	& 20.23 			& 300	 & 11.5	    &	$60\,\mathrm{s}$ \\
\enddata

\end{deluxetable*}

\begin{figure}[!ht]
\epsscale{1.15}
 \plotone{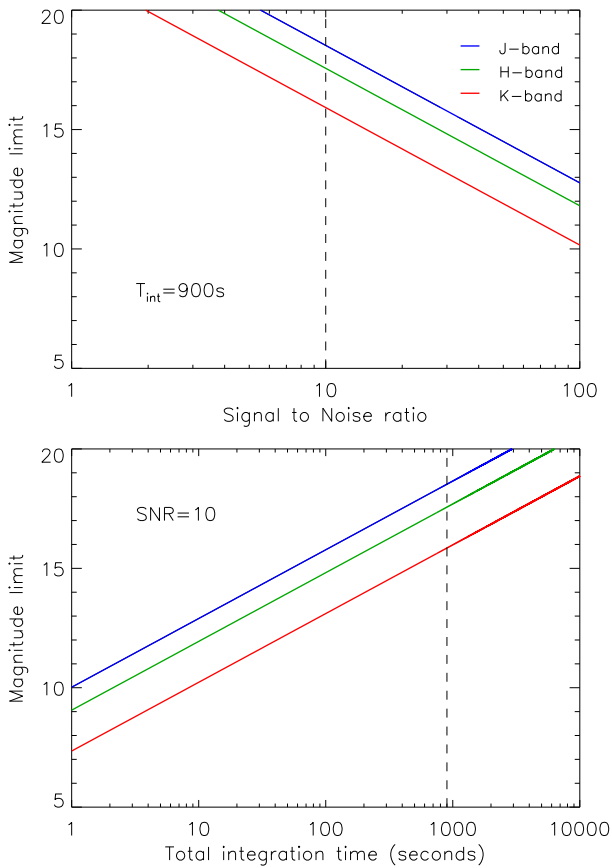} 
 \caption{Plot of estimated limiting magnitude values as a function of S/N and integration time. The \textit{top panel} shows limiting magnitude varying for different values of S/N at a fixed integration time of $900\,\mathrm{s}$. The dashed line represents a S/N of 10. Similarly, the \textit{bottom panel} shows limiting magnitude varying for different values of integration time at a S/N of 10, with a dashed line representing integration time of $900\,\mathrm{s}$.}
   \label{fig11}
\end{figure}

\begin{figure*}[!ht]
\epsscale{1.15}
 \plotone{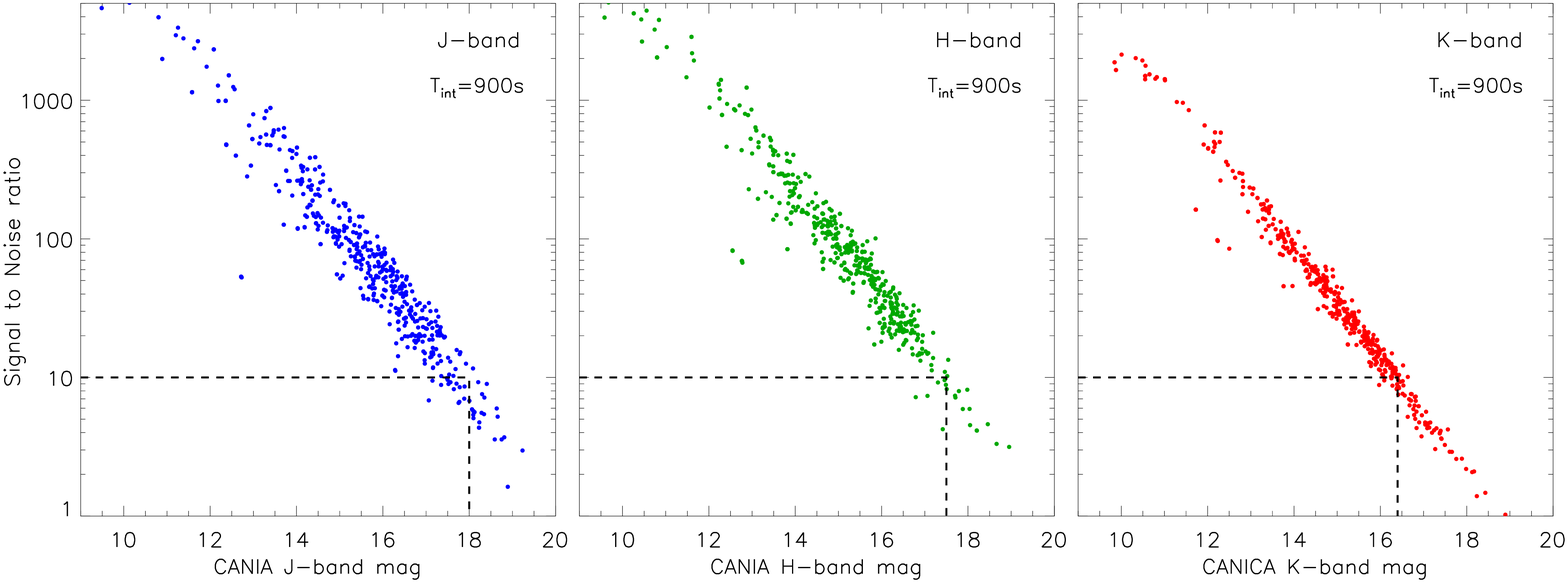} 
 \caption{Plot of S/N against magnitude for sources in AS40 field obtained in all the primary broadbands. The total integration time was fixed to $900\,\mathrm{s}$ in all the broadbands. The exposure time used in $J$, $H$, and $K$ corresponded to $60\,\mathrm{s}$, $60\,\mathrm{s}$, and $36\,\mathrm{s}$ with number of dithered images as 15, 15, and 25, respectively. The dashed line in each plot shows the S/N of 10. The magnitudes measured for this condition are $J$ = 18.0, $H$ = 17.5, and $K$ = 16.4.}
   \label{fig12}
\end{figure*}

\subsection{Limiting magnitude}

CANICA with the OAGH telescope is able to detect sources few magnitudes deeper than in the 2MASS data. Analyzing images of various standard star fields over many observing runs, we obtained the average background sky counts for a CDS image to be around $J$ = 20, $H$ = 60, and $K$ = 300 ADUs per second per pixel. This translated into sky magnitude of $J$ = 14.7, $H$ = 13.6, and $K$ = 11.5 mag per second per arcsec$^2$. The sky magnitudes set the limit of the maximum exposure for a single image before it saturates. The values of maximum exposure time in each filter are $J$ = $900\,\mathrm{s}$, $H$ = $300\,\mathrm{s}$, and $K$ = $60\,\mathrm{s}$.

The photometric magnitude for a background-limited condition can be estimated from the equation as described in \citet[see \S~9.9]{mclean08}:
\begin{equation}
Mag_{lim} = m_{zp} - 2.5alog(\frac{S/N}{g}\sqrt{\frac{N_{pix}B}{T_{int}}})
\end{equation}

where\\
$m_{zp}$ - is the zeropoint magnitude,\\
$S/N$ - is the signal-to-noise ratio,\\
$g$ - is the conversion gain in $e^{-}/ADU$,\\
$N_{pix}$ - is the number of pixel covered by a point source = $\pi(seeing/plate$-$scale)$,\\
$T_{int}$ - is the total on-source integration time, and\\
$B$ - is the background sky level in $e^{-}/s/pixel$.\\

From the measurements of zeropoint magnitudes and background values, it is estimated that with an atmospheric seeing of $1.{\arcsec}0$, CANICA is able to achieve a S/N of 10 for magnitudes $J$ = 18.5, $H$ = 17.5, and $K$ = 16.0 with an integration time of $900\,\mathrm{s}$. These values are bettered to $J$ = 20.2, $H$ = 19.3 and $K$ = 17.6 for an integration time of $1\,\mathrm{hour}$. The plots for limiting magnitude as a function of S/N and integration time are shown in Figure~\ref{fig11}. 

To verify the estimated limiting magnitudes, we observed the standard field AS40 \citep{hunt98} in all the primary broadbands. The exposure times and number of dither images were chosen to reach a total integration time of $900\,\mathrm{s}$. The exposure time values in $J$, $H$, and $K$ corresponded to $60\,\mathrm{s}$, $60\,\mathrm{s}$, and $36\,\mathrm{s}$ with number of dithered images as 15, 15, and 25, respectively. Image reduction was carried out as described in Section~\ref{obsch}. The reduced images were astrometry corrected and photometry was performed on all the point sources fixing an aperture radius of 10~pixels. The instrumental magnitudes were corrected using 2MASS magnitudes (as described in Section~\ref{photcal}) to obtain the zeropoint corrected magnitudes. Next, for each source in the field, its flux and flux error were obtained using \textit{aper} routine in IDL. The S/N was then calculated as the ratio of flux and flux error values. Figure~\ref{fig12} shows the plot of S/N against magnitude for all the sources. It is seen that for a S/N of 10, CANICA is able to reach magnitudes of $J$ = 18.0, $H$ = 17.5, and $K$ = 16.4. The observed magnitudes have differences less than $0.5\,\mathrm{mag}$ when compared with the estimated magnitudes. The relatively small differences may be due to changes in seeing, background sky variations, and photometric errors. Overall, the estimated limiting magnitudes are consistent and demonstrate the photometric capabilities of CANICA. The combined photometric performance of various parameters in all the three broadband filters are summarized in Table~\ref{tbl-3}.

\section{Summary}

We have presented characterization and performance evaluation of the Cananea Near-infrared camera at the $2.1\,\mathrm{m}$ OAGH telescope. We obtained the key detector parameters of conversion gain, dark current, readout noise, and linearity. Gain measurement of the detector was performed using the photon transfer curve by estimating the reciprocal of the slope of a variance-signal plot, which returned a value of $5.8\,\mathrm{e^-/ADU}$. The dark current values were obtained from a series of darks taken at different exposure times. The time-dependent dark current is estimated to be $1.2\,\mathrm{e^-/sec}$. Readout noise was obtained using a large sample of BIAS images associated to the dark exposures. The average readout noise for CDS technique was measured to be $30\,\mathrm{ e^-/pixel}$. The HAWAII detector array showed non-linearity up to 10\% close to the full-well-depth. We developed a simple and fast technique to perform linearity correction, which reduced the non-linearity to under 1\% levels for CDS images.

The CANICA observing scheme uses a dithering methodology to acquire 15 images in a single sequence. The dithered images are reduced and analyzed by a customized pipeline developed in IRAF. The main reduction steps involved in linearity corrected images are dark subtraction, flat fielding and sky subtraction. Flat fielding is performed using dome flats obtained with lights ON and OFF technique. Sky image is obtained by taking median of all the dithered images. Astrometry corrections and photometric calibration are performed via comparison to the publicly available 2MASS data. The zeropoint magnitudes for each observing field are measured from the average of the differences of 2MASS and instrumental magnitudes of the sources.

Imaging performance of CANICA was evaluated by measuring the PSF, zeropoints, throughputs, and limiting magnitude. Full-field analysis of the PSF showed less than 10\% variations in measured FWHM values across the central $4 \times 4\,\mathrm{arcmin^2}$. The average zeropoint values in the primary broadbands are calculated to be $J$ = 20.52, $H$ = 20.63, and $K$ = 20.23. The zeropoint values across the FOV showed variations less than $0.1\,\mathrm{mag}$, implying that the changes in PSF across FOV do not affect photometric measurements. The total throughput of all systems involved with CANICA was measured to be $J$ = 9.48\%, $H$ = 11.26\%, and $K$ = 14.84\%. The background-limited magnitudes reached by CANICA on the OAGH telescope are $J$ = 18.5, $H$ = 17.6, and $K$ = 16.0 for a S//N of 10 with an integration time of $900\,\mathrm{s}$.

CANICA with a FOV of $5.5 \times 5.5\,\mathrm{arcmin^2}$ and a plate scale of $0.32\,\mathrm{arcsec/pixel}$ offers highly effective imaging and photometric capabilities. The results obtained from evaluation of CANICA will help to conduct customized observations for studying various astrophysical topics in the NIR. The characterization of CANICA and photometric calibration presented here sets the base for polarimetric observations using the recently commissioned instrument: POLICAN. The detailed description of methods used in the article will help in characterizing and evaluating performance of similar NIR cameras.

\acknowledgements

We thank the anonymous referee for the useful comments. The calibration observations were performed with the help of OAGH staff; we thank them for their support and technical help with the instrument. We thank Dan Clemens, Boston University, for providing valuable comments on improving the article. We would also like to thank Massimo Robberto, STScI, for helpful discussion on linearity corrections. This work was been carried out at Instituto Nacional de Astrof\'isica, \'Optica y Electr\'onica, M\'exico with support from CONACyT projects CB-2002-01 G28586E (P.I. LC), CB-2010-01 155142-43 (P.I. Y.D.M.) and CB-2012-01 182841 (P.I. AL). R.D. with CVU 555629 acknowledges CONACyT for the grant 370405. This work makes use of data products from the Two Micron All Sky Survey, which is a joint project of the University of Massachusetts and the Infrared Processing and Analysis Center/California Institute of Technology, funded by the National Aeronautics and Space Administration and the National Science Foundation. 

\appendix
\section{Error propagation in calculation of conversion gain} \label{appA}
The conversion gain is measured by estimating the reciprocal of the slope of a variance-signal plot obtained from a series of dome illuminated flats. The pixel-to-pixel variations in individual flats are corrected by dividing with a normalized flat. The error propagation after correction of individual flat images can be given as 
\begin{equation}\label{eqnB1}
(\frac{\sigma_{corr}}{S_{corr}})^2 = (\frac{\sigma_{ind}}{S_{ind}})^2 + (\frac{\sigma_{flat}}{S_{flat}})^2
\end{equation}
where $S_{corr}$,$S_{ind}$,$S_{flat}$ and $\sigma_{corr}$,$\sigma_{ind}$,$\sigma_{flat}$ are signal and standard deviation of the corrected, individual, and normalized flat images, respectively. The variance of each image is the square of its standard deviation.

The total signal ($S_{flat}$) of a normalized flat is equal to $\sim1.0$, as normalization is done by using the mean of the pixel values. Also the noise (standard deviation) in the normalized flat image is same as the noise in individual mean flat image (i.e.\ $\sigma_{flat} = \sigma_{ind}$), as both are obtained from same set of dome flats. Now we can rewrite the equation \ref{eqnB1} as
\begin{equation}
(\frac{\sigma_{corr}}{S_{corr}})^2 = (\frac{\sigma_{ind}}{S_{ind}})^2 + (\frac{\sigma_{ind}}{1})^2
\end{equation}

This equation can be rearranged and written as

\begin{equation}
(\frac{\sigma_{corr}}{\sigma_{ind}})^2 = (\frac{S_{corr}}{S_{ind}})^2 (1+S_{ind}^2)
\end{equation}

The above equation gives the ratio of variance of corrected and individual flat. Hence, the terms in the right-hand side represent the change in variance due to flat fielding, which is the increase in error.

\section{Error propagation in measurement of readout noise} \label{appB}
During the measurement of readout noise, the individual BIAS images are differenced with the mean BIAS image (see Section~\ref{readn}). The noise (standard deviation) in the differenced image ($\sigma_{diff}$) can be described by standard error propagation as follows:
\begin{equation} \label{eqnA1}
\sigma_{diff} = \sqrt{\sigma_{ind}^2 + \sigma_{avg}^2}
\end{equation}
where $\sigma_{ind}$ is the noise in individual BIAS image and $\sigma_{avg}$ is the noise in mean BIAS image. Because the mean BIAS image is constructed from ``N'' number of individual BIAS images, $\sigma_{avg}$ can be related to $\sigma_{ind}$ as
\begin{equation}
\sigma_{avg} = \frac{\sqrt{N \sigma_{ind}^2}}{N}
\end{equation}

Because we use 10 BIAS images for obtaining the mean image, we can write the noise in mean BIAS image as:
\begin{equation}
\sigma_{avg} = \frac{\sqrt{10 \sigma_{ind}^2}} {10} = \frac{\sigma_{ind}}{\sqrt{10}}
\end{equation}

Substituting the above value in equation \ref{eqnA1}, we get
\begin{equation}
\sigma_{diff} = \sqrt{\sigma_{ind}^2 + \frac{\sigma_{ind}^2}{10}} = \sqrt{1.1 \sigma_{ind}^2}
\end{equation}

We can rewrite the above equation as
\begin{equation}
\frac{\sigma_{diff}}{\sigma_{ind}} = \sqrt{1.1}
\end{equation}

The above equation gives the ratio of noise values of differenced and individual BIAS images. Hence, the term in the right-hand side represents the change in noise measurements due to the differencing of mean BIAS image with the individual BIAS images.



\end{document}